\documentclass[12pt]{iopart}
\usepackage{iopams}
\usepackage{graphicx}

\begin{document}
\title[BW null-model]{Selective pressure on metabolic network structures as measured from the random blind-watchmaker network.}
\author{Sebastian Bernhardsson and Petter Minnhagen}
\address{IceLab, Dept. of Physics, Ume\aa \ University. 901 87 Ume\aa . Sweden}
\ead{sebbeb@tp.umu.se}


\begin{abstract}
A random null model termed the \emph{Blind Watchmaker network} (BW) has been shown to reproduce the degree distribution found in metabolic networks. This might suggest that natural selection has had little influence on this particular network property. We here investigate to what extent other structural network properties have evolved under selective pressure from the corresponding ones of the random null model: The clustering coefficient and the assortativity measures are chosen and it is found that these measures for the metabolic network structure are close enough to the BW-network so as to fit inside its reachable random phase space. It is furthermore shown that the use of this null model indicates an evolutionary pressure towards low assortativity and that this pressure is stronger for larger networks. It is also shown that selecting for BW networks with low assortativity causes the BW degree distribution to slightly deviate from its power-law shape in the same way as the metabolic networks. This implies that an equilibrium model with fluctuating degree distribution is more suitable as a null model, when identifying selective pressures, than a randomized counterpart with fixed degree sequence, since the overall degree sequence itself can change under selective pressure on other global network properties.
\end{abstract}

\maketitle

\section{Introduction}

A network is a representation of whom or what is connected to, or influenced
by, whom or what. To characterize the structure of a network, one
often starts with the degree distribution $N(k)$, the number of nodes with $k$
connections. Numerous studies have shown that real-world networks often
have very broad degree distributions for larger $k$, $N(k)\sim k^{-\gamma}$%
\cite{albert02}\cite{dorog03}\cite{newman03}\cite{bocca06}\cite{newman06}.
Biological networks are particularly interesting because the structure of these networks have,
directly or indirectly, arisen through the process of evolution by natural
selection. These networks have been metaphorically constructed as if by a blind watchmaker,
through the interplay between a random stochastic evolution and a
selection process \cite{dawkins}. The scale-free
structure of biological networks are often ascribed to various aspects of the evolutionary
process: Either the scale-free network structure has been suggested to
confer an evolutionary advantage\cite{albert00}\cite{cohen00}, or the
elementary mechanism for growth of the network has been suggested to
generate \emph{a priori} a scale-free network \cite{barabasi99}\cite{sole02}
\cite{vazquez03}\cite{Pfeiffer}. It has also been shown that the outcome of many dynamical processes on biological networks are strongly coupled to the overall network structure \cite{Li}\cite{Wardil}.

It has recently been shown that the degree distribution of metabolic networks
can to a large extent be reproduced by a maximum entropy solution \cite{Minnhagen07} of a random network model,
 called the \emph{blind watchmaker network} (BW) \cite{bernhardsson08}.
This finding implies that evolutionary pressure has had little or no influence on the degree distribution,
since there is no or very little deviation from the random null model which is presumed to describe the resulting structure in the absence of any selection. 
We use the term \emph{null model} following the motivation by Balcan et al. in Ref.\ \cite{balcan04} that such a model is void of any assumptions for any type of fitness of any type of interaction on which natural selection can subsequently act. Nevertheless, the metabolism in itself is a fine tuned machinery which has indeed been shaped by natural selection. 
If a specific network quantity is constrained and yet the observed degree distribution is close to the one corresponding to the constraint-free maximization of entropy, then the selective pressure associated with the constrained quantity has obviously little or no influence on the degree-distribution. In such a case the non-random network structure associated with the constraint should fit inside the phase space spanned by the blind watchmaker network.
Also, structures that deviate from this random null model signals a directional pressure and thus gives us information about which network properties are selected for. The data used for the real metabolic networks is taken from references \cite{ma03a,ma03b} which contains a full set of known metabolic reactions and the corresponding substances involved for 107 different organisms. We use the undirected substance network-representation which means that two substrates are connected if one is used in a reaction to create the other.

Here we investigate the clustering-assortativity space of the blind watchmaker network, in a similar way as in Ref.\ \cite{holme07}, and compare to metabolic networks. However, there is an important difference to the approach in Ref.\ \cite{holme07}: This earlier work explored the possible phase space of a given, fixed, degree distribution. In the present work we have an open system where also the degree distribution is free to adjust, because our null model is defined by a random process. Thus, we can obtain a network with a prescribed assortativity (or some other network structure measure) by just selecting those networks from the ensemble of networks generated by the random process. This is an appealing feature, since it does not require any \emph{ad hoc} assumption about the time evolution like e.g. the preferential attachment scheme. 

\section{Results}
The Blind  Watchmaker network is the null model for a network of which one only has limited knowledge \cite{bernhardsson08}: it is the most likely network structure for the given limited information. This network structure is obtained from variational calculus as the maximum entropy solution where the limited information enters as constraints \cite{bernhardsson08}. The constraints for the BW-network are in general the total number of nodes $N$ and the total number of links $M$ together with the usual network constraints (no self-loops and no multiple links). In addition it is assumed that there is neither an \emph{a priori} preference as to which links are joined nor in which order.  This lack of \emph{a priori} preference defines the \emph{a priori}  randomness inherent in the BW-network. 

A convenient way of obtaining the variational solution is by devising a numerical algorithm which operates under the same randomness and constraints and hence converges to the same solution.  However, it is important to realize that this numerical algorithm has nothing \emph{per se} to do with the any actual network evolution, but is just a numerical device to obtain the correct network structure.

An algorithm for generating the Blind Watchmaker networks is described in Ref.\ \cite{bernhardsson08}. It encodes equal probability for a link-end to rewire to the same node as an arbitrarily chosen other link-end, as well as, equal probability for the relative order in which two link-ends arrive at the same node. This can be implemented as follows:
Start from a set of links attached to a set of nodes in an arbitrary way \cite{bernhardsson08}.
\begin{itemize}
 \item[1] Pick two nodes, A and B, randomly each with a probability proportional to the square of their degree, $p_i\propto k_i^{2}$.
 \item[2] Pick a random link on node A and move it to node B.
 \item[3] If the attempted move is forbidden by a constraint choose another link from the same node
A and repeat until a link has been moved or until all links have been proven to violate the constraint.
\end{itemize}

The default constraints are 1) no self loop, 2) no multiple links between nodes, 3) no zero degree nodes. When repeated until a steady state has been reached the algorithm creates a scale-free network with an exponent around 2. This is illustrated in Fig.\ 1(a) where the average result is shown for 107 BW networks together with 107 metabolic networks \cite{ma03a,ma03b}. In the special case of metabolic networks there is an additional, presumably chemical, constraint\cite{footnote} on the number of one degree nodes (nodes with only one link). The BW-network including this additional substance-constraint is also shown in Fig.\ 1(b). In this figure the expectancy values are shown (black curve), instead of a scatter plot, together with the curves corresponding to two standard deviations away from the average solution, illustrating the spread generated by the BW-algorithm. The expectancy values and the corresponding spread was calculated from 1000 sets of networks, each containing 107 independent samples.
In this comparison the BW-networks in Fig.\ 1(a) have the same average size and the same average number of links per node as the metabolic ensemble and in Fig.\ 1(b) also the same average number of nodes with one link. The agreement in Fig.\ 1(b) is particularly striking. 

\begin{figure}[!tp]
\begin{center}
\includegraphics{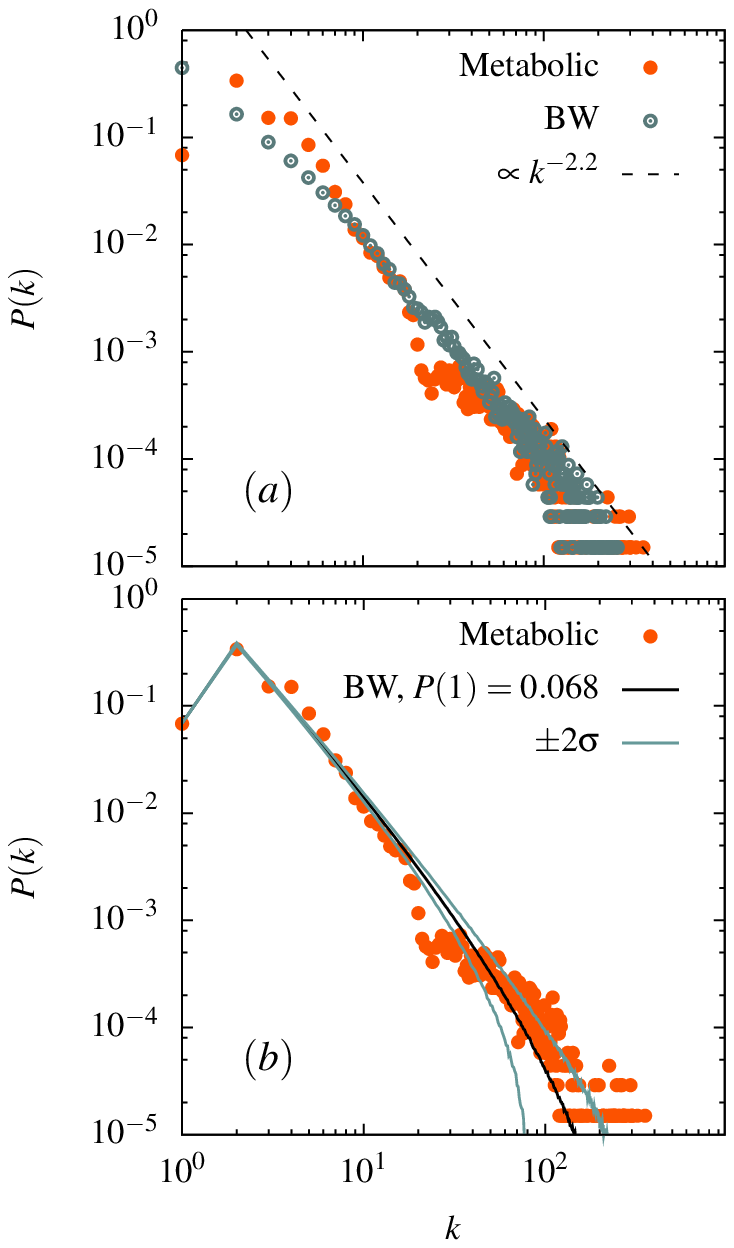}
\end{center}
\caption{Degree distribution of 107 metabolic networks (triangles) and 107 BW networks in log-log scale (a) without constraints, and (b) with a constraint on the number of one degree nodes, $P(1)$.
Figure (a) shows the average solution of 107 BW samples, while (b) shows the expectancy values (black curve) plus the spread (blue curves showing $\pm$ two standard deviations).}
\label{BW_vs_metab}
\end{figure}

In order to further quantify the network structure we choose the clustering coefficient and the assortativity. Both measures are represented by a single real number and we can thus further characterize a network by a point in a clustering-assortativity space.
The clustering coefficient (CC) is a measure of the number of triangles existing in a network, normalized by the possible number of triangles that could exist,

\begin{equation}
C_i = \frac{2N_{\bigtriangleup}}{k_i(k_i-1)},
\label{local_clust_coef}
\end{equation}
where $N_{\bigtriangleup}$ is the number of triangles (three nodes where everyone is connected to everyone) and $k_i$ is the degree of node $i$. A total average CC can then be calculated as 
\begin{equation}
C = \frac{1}{N_{k>1}}\sum_{i,k_i>1} C_i,
\label{tot_local_clust_coef}
\end{equation}
where $N_{k>1}$ is the number of nodes with a degree larger than one. 

The second measure, the assortativity, is based on the Pearson correlation coefficient which ranges between the values -1 and 1. It is defined for networks as \cite{newman02}
\begin{equation}
r = \frac{4\langle jk \rangle - \langle j+k \rangle^2}{2\langle j^2+k^2 \rangle-\langle j+k \rangle^2},
\label{assort}
\end{equation}
where $\langle ... \rangle$ means an average over all links and $j$ and $k$ are the degrees of the nodes on either side of a link.
$r=-1$ means perfect disassortative mixing (connected nodes have very different degrees) and $1$ means perfect assortative mixing (connected nodes have the same degree).

\begin{figure}[!t]
\begin{center}
\includegraphics{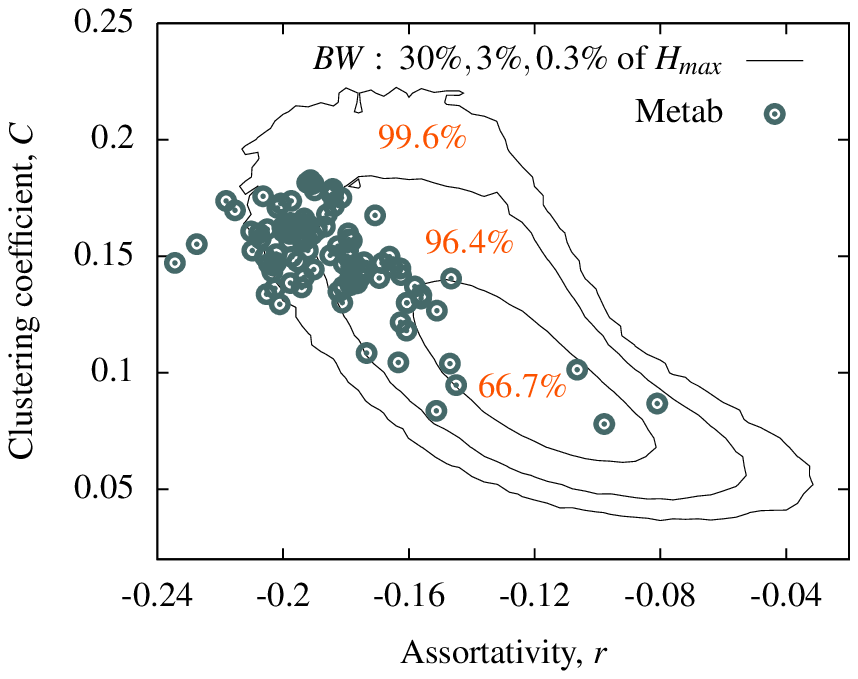}
\end{center}
\caption{The clustering-assortativity space. The points represent 107 different metabolic networks and the contour lines corresponds to $30\%$, $3\%$ and $0.3\%$ of the maximum density of BW networks. The contour lines then enclose $68\%$, $96\%$ and $99.6\%$, respectively, of all the generated BW networks with $N=640$ and $\langle k \rangle = 5.35$.}
\label{Cr}
\end{figure}

The phase space of the BW model is investigated by generating many ($\sim10^6$) networks and measuring their CC and assortativity. The C-r space is then discretized and a two dimensional histogram, representing the density of occurrences, is created for each interval in C and r. The result is then plotted as contour lines corresponding to $30\%$, $3\%$ and $0.3\%$ of the maximum height, $H_{max}$, of the histogram surface, and thus enclosing (as measured from the generated data) $66.7\%$, $96.4\%$ and $99.6\%$ of all the generated networks. The contours are chosen so as to correspond to a standard two dimensional Gaussian contour plot for $1$, $2$ and $3$ standard deviations away from the mean.
Figure \ref{Cr} shows the result for an ensemble of BW networks of the same size as the average metabolic network with $N=640$ and $\langle k \rangle = 5.35$. The points represent the metabolic networks (one point per network) which are included for comparison.
The first impression of the figure is perhaps the clear correlation between the values of $C$ and $r$, which seems to be shared by the real networks. Low assortativity is coupled with high clustering. It is interesting to note that this result is opposite to what was found in Ref.\ \cite{holme07} which considered a system with a fixed degree sequence. In that case low assortativity was coupled with low clustering.
Another interesting feature is that the BW model includes a wide range of C and r values. The overlap of the BW phase space and the metabolic networks suggests that these structural features are not completely random. In addition it shows that these non-random features can be obtained directly from the BW process by selection. This suggests a selective pressure towards lower assortativity and higher clustering for the metabolism of most organisms.

\begin{figure}[!tp]
\begin{center}
\includegraphics{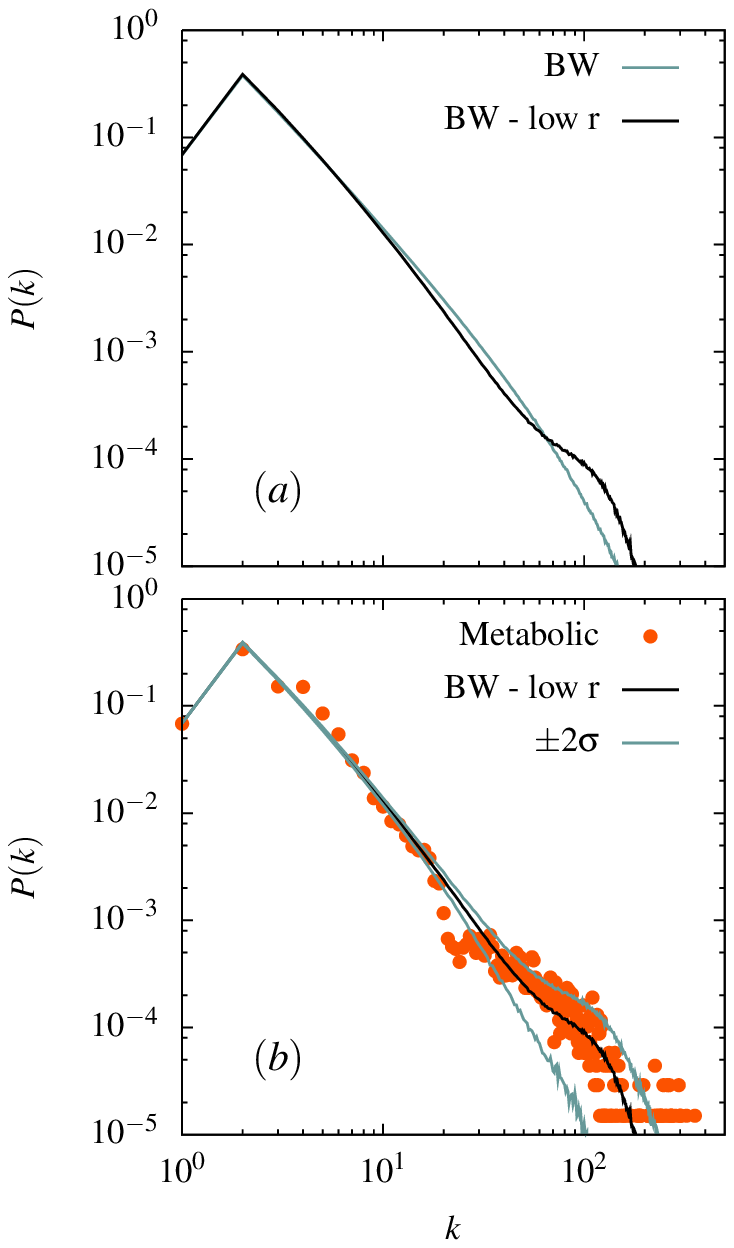}
\end{center}
\caption{(a) A comparison between the original BW model (blue curve) and BW networks with a low assortativity in the range $-0.21 < r < -0.19$ (black curve).
(b) The degree distribution for the metabolic networks which have an assortativity in the range $-0.21 < r < -0.18$ (circles), and the BW networks with low assortativity, giving the same $\langle r \rangle$ for both systems. The spread of the BW process is shown by the blue curves ($\pm$ two standard deviations). Both figures are plotted in log-log scale.}
\label{low_r}
\end{figure}

What happens with the degree distribution when selecting for lower assortativity? If the similarity between the degree distributions becomes dramatically worse then the BW null model is in fact inconsistent with the metabolic networks. 
In order to investigate how the selection affects the BW degree distribution we pick out realizations within a prescribed narrow assortativity interval. In Fig.\ \ref{low_r} we show the average result of 62 BW networks with an assortativity in the range $-0.21 < r < -0.19$ with an average r of $\langle r \rangle = -0.195$ together with the 62 metabolic networks that lay in the range $-0.21 < r < -0.18$ with approximately the same average $r$. Both the BW networks and the average of these metabolic networks have the size $N=730$ and $\langle k \rangle = 5.5$.
As seen from the Figs 3a and b, the BW degree distribution changes slightly in the tail part. More accurately, Fig.\ 3a shows that compared to the original BW distribution, the selection for low r creates a dip for intermediate degrees and a bump for high degrees. In fact, Fig.\ 3b shows that this change makes the BW distribution even more similar to that of the metabolic networks. This suggests that the selection for low assortativity is in fact in this case a selection \emph{of} a degree-distribution and not a selection \emph{within} a given fixed degree-distribution. 

This is further illustrated in Fig.\ 4(a) and (b) where the data is plotted in the same contour plot as in Fig.\ 2. Figure 4(a) shows the overlap between the BW phase space and 62 metabolic networks with low $r$ (circles). The triangles show the result of randomizing each of the 62 metabolic networks keeping all nodes individual degree \cite{maslov02} and averaged over 100 independent randomizations. The small shift between the clouds of circles and triangles again indicates that much of the network structure (represented by the clustering and assortativity) is absorbed into the degree sequence.
Figure 4(b) shows the same thing as in (a) but for 62 selected snapshots of the BW networks lying in the small $r$ range. These networks has also been randomized keeping the degree sequence (triangles) in the same way as in (a) and the BW null model seem to behave very similar to the metabolic networks.
Furthermore, the CC for these selected BW networks is $\langle C \rangle = 0.17$ compared to $\langle C \rangle = 0.16$ for the corresponding 62 metabolic networks. This means that, when selecting for a low assortativity, the BW networks automatically obtain a high clustering.

\begin{figure}[!tp]
\begin{center}
\includegraphics{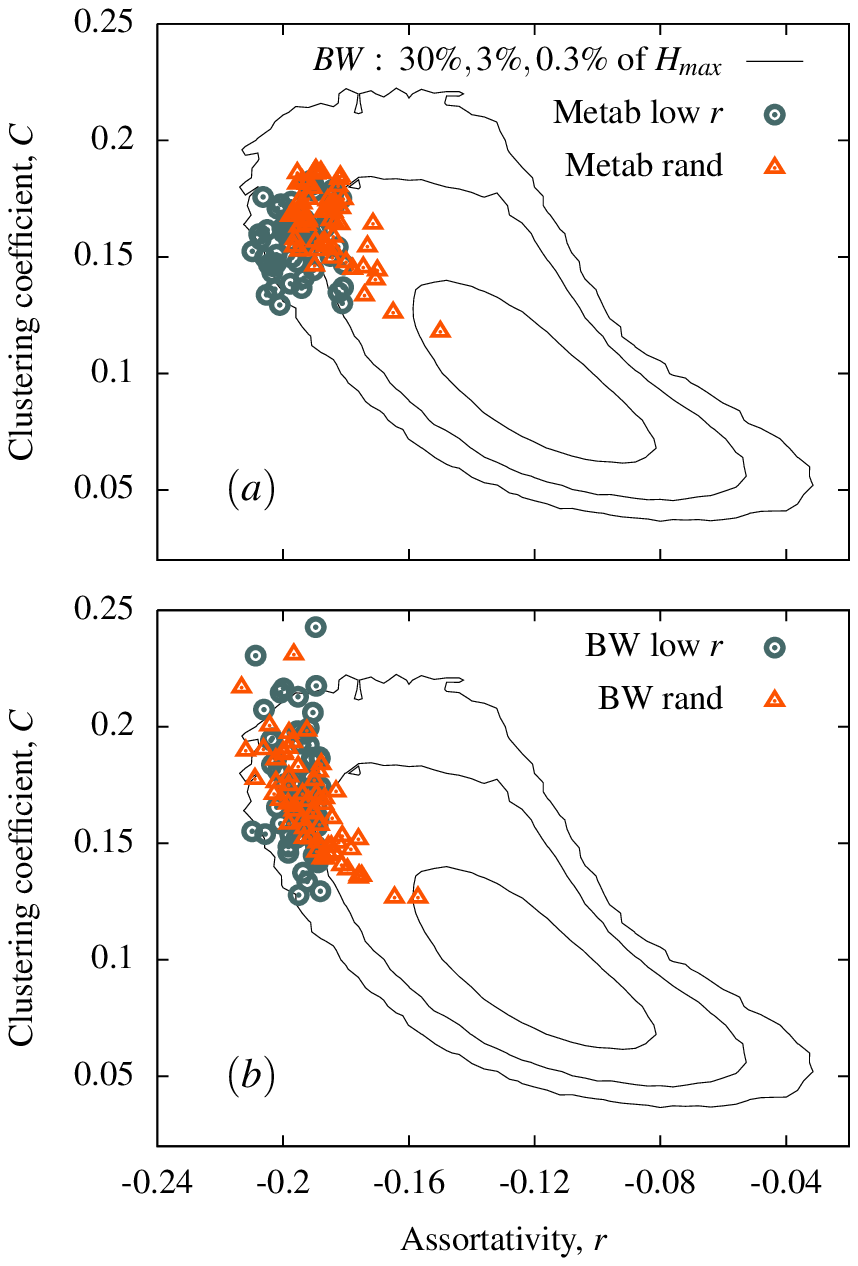}
\end{center}
\caption{Low assortativity: (a) 62 metabolic networks with low assortativity (see Fig.\ 3) as circles and their randomized counterparts (keeping the degree sequence) as triangles. (b) Same thing for 62 selected BW networks in the low $r$ range.}
\label{Cr_low_r}
\end{figure}

\begin{figure}[!tp]
\begin{center}
\includegraphics[width=\columnwidth]{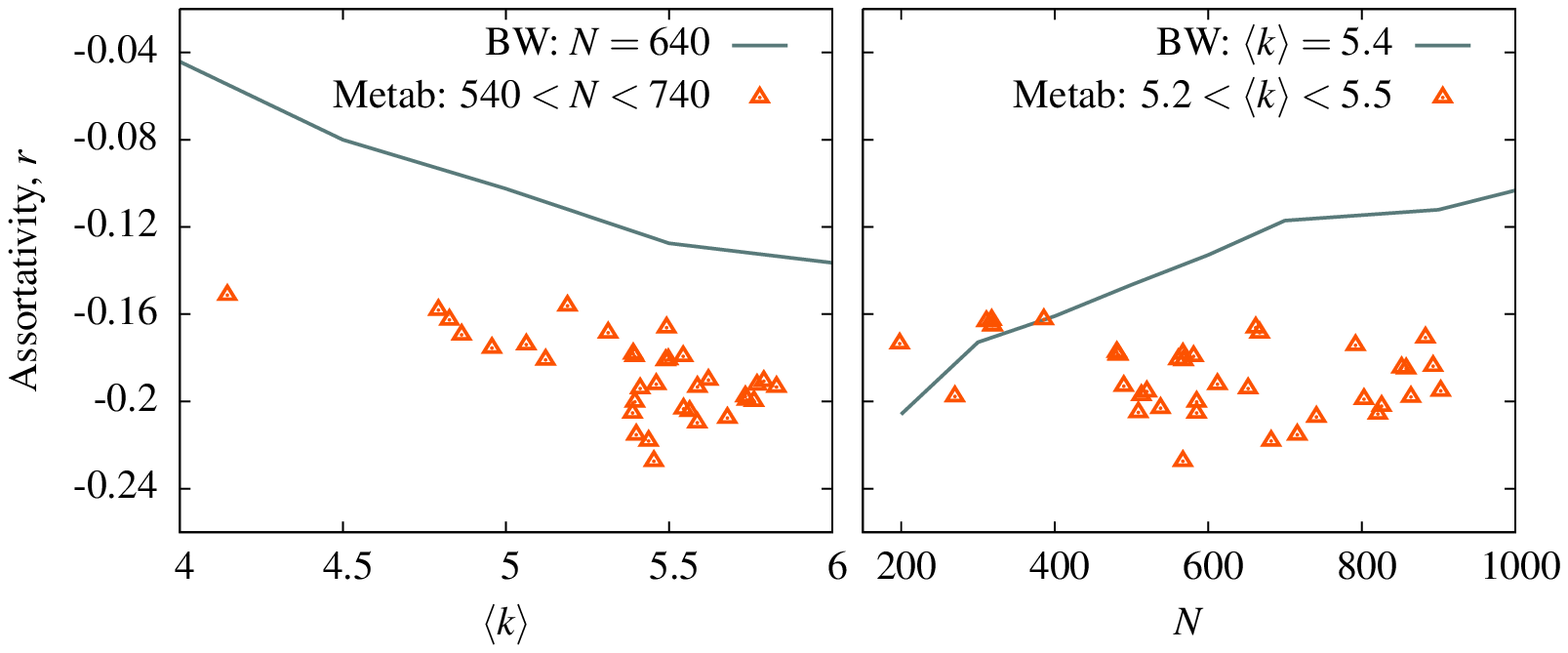}
\end{center}
\caption{The dependency of the assortativity ($r$) on the average degree ($\langle k \rangle$) and the number of nodes ($N$) for the BW null model (line) and the metabolic networks (triangles): (a) $r$ as a function of $\langle k \rangle$ for $N=640$ for the BW networks and $540 < N < 740$ for the metabolic networks. (b) $r$ as a function of $N$ for $\langle k \rangle=5.4$ for the BW networks and $5.2 < \langle k \rangle < 5.5$ for the metabolic networks.}
\label{sizedep}
\end{figure}

Metabolic networks of different organisms have very different sizes and average degrees. The number of nodes varies from about 200 for the smallest networks to almost 1200 for the largest, and the smallest average degree is around 4 while the largest is close to 6. This fact has not been taken into account in the analysis of Fig.\ \ref{Cr}-\ref{Cr_low_r}. To study the effect on the structural measures, both for the BW null model and the real metabolic networks, we plot the assortativity as a function of $N$ (Fig.\ \ref{sizedep}(a)) and as a function of $\langle k \rangle$ (Fig.\ \ref{sizedep}(b)). We only investigate the size dependency on the assortativity, since this measure is highly correlated with the clustering coefficient, and will thus give a representative behavior for both measures. Since the metabolic networks possess a wide spread in both $N$ and $\langle k \rangle$, a smaller sampling interval is used for one of them, while sampling a broader interval for the other. In Fig.\ \ref{sizedep}(a) the assortativity is plotted as a function of the average degree for the BW networks with $N=640$ together with the metabolic networks in the range $540 < N < 740$. Both the BW and the metabolic networks display a decrease in the assortativity, when the average degree is increased with almost a constant shift between them. This constant shift again signals a selective pressure towards lower assortativity, which is independent of the average number of links in the system. Figure \ref{sizedep}(b), on the other hand, is showing an increase in the assortativity for the BW networks as a function of the number of nodes, while for the metabolic networks it is basically independent of the size. Here, $\langle k \rangle = 5.4$ is used for the BW networks. In this case the metabolic networks with an average degree in the range $5.2 < \langle k \rangle < 5.5$ show no increase with $N$. This means that the difference between the BW-network and the metabolic network increases with $N$. Thus the results presented in Fig.\ \ref{sizedep}(b) suggest that there is essentially very little selective pressure on the assortativity for small metabolic networks and that this pressure is increases with the network size.

\section{Discussion}
In this work we investigate the clustering-assortativity phase space of the Blind Watchmaker(BW) network model and compare to the real data of 107 metabolic networks. We show that the structural network properties captured by these two network measures are directly reachable within the BW null model without any assumptions about the evolutionary path. It is also demonstrated that when selecting for the BW networks that posses the same structural properties as the real data the resulting degree distribution is effected. Furthermore, the direction of this change appears to be towards increased similarity. This implies a coupling between the degree distribution and other structural properties like assortativity and clustering. Thus the small deviation between the degree distributions of the metabolic networks and the BW null model is according to our analysis suggested to be caused by a selection towards lower assortativity.

We also find that the clustering- and assortativity measures are correlated in the same way both for the BW model and for the real metabolic networks. It was noted that this correlation is the reverse of what was found by in Ref.\ \cite{holme07} when mapping out the possible phase space for a given, \emph{fixed}, degree sequence. Our results implies that the structural properties of a network can depend on the degree distribution in a crucial way, which limits the usefulness of drawing conclusions on structural interdependencies from fixed degree distributions. In this respect a random null model, like the BW-model where degree-distribution is allowed to fluctuate, gives a better starting point when trying to identify selection pressures. This point is further clarified by randomizing the metabolic networks using the Maslov-Sneppen \cite{maslov02} routine which preserves the individual degrees. This randomization changes the average assortativity by only $3.3\%$, suggesting that the metabolic networks are random and void of any significant selection for assortativity. However, our comparison to the BW networks gives a much larger difference ($33\%$), because the selection really causes a \emph{change} in the actual degree sequence. This implies that it is important to identify an adequate null model which does not have a fixed degree sequence, when searching for selective pressure in networks. We suggest that in this respect the Blind Watchmaker network is an appropriate null model for metabolic networks.

One should however note that a null model like the Blind Watchmaker network does not \emph{per se} give any information of the precise metabolic evolution: it is just the best guess of the structure you can make \emph{provided} you have no knowledge of the evolutionary process, except what is imposed by global constraints. From this point of view it is the deviation between the null model and the actual metabolic network which contains the most interesting information. The fact that this deviation is small suggests that whatever the explicit metabolic evolution path might have been, it has had surprisingly little influence on the global metabolic network structure. Our results also imply that \emph{a priori} identification of additional relevant global constraints for the BW-model is one possible way of gaining further understanding of the overall metabolic network structure.


\begin{thebibliography}{99}
\bibitem{footnote} The number of enzymatic reactions with either only one substrate
or one product and the number of substances which take part in only one reaction is relatively small for metabolic networks.
\end{thebibliography}
\end{document}